\documentclass{sig-alternate-10pt}

\PassOptionsToPackage{hyphens}{url}
\usepackage[hyphens]{url}
\usepackage{hyperref}
\hypersetup{breaklinks=true}
\usepackage{balance}

\usepackage{algorithm}
\usepackage{algorithmic}
\usepackage{amsfonts}
\usepackage{amssymb}
\usepackage{boxedminipage}
\usepackage{caption}
\usepackage[compress,sort]{cite}
\usepackage{color}
\usepackage{colortbl}
\usepackage{enumerate}
\usepackage{epsfig}
\usepackage{graphicx}
\usepackage{lastpage}
\usepackage{listings}
\usepackage{multirow}
\usepackage{paralist}
\usepackage{pifont}
\usepackage{rotating}
\usepackage{tabularx}
\usepackage[usenames,dvipsnames]{xcolor}
\usepackage{xspace}
\usepackage{comment}
\usepackage{subfigure}
\usepackage{fancyvrb}

\usepackage{enumitem}

\includecomment{notready}
\excludecomment{ready}
\begin{ready}
  \newcommand{\plonka}[1]{\typeout{\the\inputlineno: PLONKA: #1}}
  \newcommand{\arthur}[1]{\typeout{\the\inputlineno: ARTHUR: #1}}
\end{ready}
\begin{notready}
  \newcommand{\plonka}[1]{{\color{red}{\it plonka - #1}}\typeout{\the\inputlineno: PLONKA: #1}}
  \newcommand{\arthur}[1]{{\color{cyan}{\it arthur - #1}}\typeout{\the\inputlineno: ARTHUR: #1}}
\end{notready}

\let\oldSim\sim
\renewcommand{\sim}{\raise.17ex\hbox{$\scriptstyle\oldSim$}}

\title{kIP: a Measured Approach to IPv6 Address Anonymization}

\numberofauthors{2}

\author{
\alignauthor David Plonka \\
   \affaddr{Akamai Technologies} \\
   \email{plonka@akamai.com}
\alignauthor Arthur Berger\\
   \affaddr{Akamai Technologies \\ 
    Massachusetts Institute of Technology} \\
   \email{arthur@akamai.com}
}

\date{}

\let\OldUrlFont\UrlFont \renewcommand{\UrlFont}{\small\OldUrlFont}

\begin{document}

\maketitle

\begin{abstract}

Privacy-minded Internet service operators anonymize IPv6 addresses by
truncating them to a fixed length, perhaps due to long-standing
use of this technique with IPv4 and a belief that it's ``good
enough.''  We claim that simple anon\-ym\-ization by truncation is
suspect since it does not entail privacy guarantees nor does it take
into account some common address assignment practices observed
today. To investigate, with standard activity logs as input, we
develop a counting method to determine a lower bound on the number
of active IPv6 addresses that are {\em simultaneously assigned,}
such as those of clients that access World-Wide Web services.
In many instances, we find that these empirical measurements offer
no evidence that truncating IPv6 addresses to a fixed number of bits,
{\em e.g.,} 48 in common practice, protects individuals' privacy.

To remedy this problem, we propose {\em kIP anonymization,} an
aggregation method that ensures a certain level of address privacy.
Our method adaptively determines variable truncation lengths using
parameter {\em k,} the desired number of {\em active} (rather
than merely {\em potential}) addresses, {\em e.g.,} 32 or 256,
that can not be distinguished from each other once anon\-ym\-ized.
We describe our implementation and present first results of its
application to millions of real IPv6 client addresses active over
a week's time, demonstrating both feasibility at large scale and
ability to automatically adapt to each network's address assignment
practice and synthesize a set of {\em anonymous aggregates}
(prefixes), each of which is guaranteed to cover (contain) at
least {\em k} of the active addresses.  Each address is anonymized
by truncating it to the length of its longest matching prefix in
that set.

\end{abstract}

\section{Motivation \& Introduction}

Protecting personally identifiable information (PII) in the form
of IP addresses warrants special attention with IPv6 due both
to nascent privacy concerns and mandates, {\em e.g.,} in the
European Union, and to increased IPv6 use, worldwide.
Given today's significant IPv6 deployment and dual-stack operation,
the IPv6 address may be the identifier most likely to be unique to a client on
the increasingly encrypted World-Wide Web (WWW). While individual
IPv4 addresses are increasingly shared due to address exhaustion,
this is neither intended nor commonplace with IPv6 which offers
unique, globally-routed addresses end-to-end.

In this work we investigate but one Internet privacy
measure: IP address anonymization by {\em truncation}. Address
truncation means simply to delete a set of contiguous low
(rightmost) bits, {\em i.e.,} to remove a suffix from an
input address. Typically the suffix' bits are replaced with zeroes
so that the anonymized output is an address-sized value. While
more complex anonymization techniques have been implemented and are
well-studied~\cite{DBLP:conf/imw/XuFAM01,DBLP:journals/cn/FanXAM04},
they anonymize addresses in a way that prevents the result from
being used for standard security, operations, and research tasks.
Specifically, they prevent correlation with network topology,
routing, service providers, and locations.  For these purposes,
trun\-cation-based anonymization is ideal {\em if, and only if,}
it can be guaranteed to improve privacy.

Such anonymization is typically performed by truncating input
addresses to one fixed length.  Consider, for instance, a WWW
analytic system employing trun\-cat\-ion\--based
IP address anonymization; {\em e.g.,} zeroing the last 8 bits of
a user's IPv4 IP address and the last 80 bits of an IPv6
address~\cite{googleAnalytics}. Essentially, this is equivalent
to masking or aggregating to /24 and /48 prefixes, respectively,
perhaps combining information about as many as 256 IPv4 addresses or
64K IPv6 /64 prefixes.
Of
course, the utilization of the IPv4 and IPv6 address spaces
differ dramatically.  While someone might believe that an IPv4 /24
prefix would aggregate individual users' addresses~\footnote{Evaluating
anonymization of IPv4 addresses by truncation is warranted as
well, but it is not the subject of this work.}, we ask two
questions. First, can passive measurements inform
decisions about anonymization? Second, is
there reason to believe that any one IPv6 prefix length would
perform satisfactorily?

The key problem is how to decide at what prefix (bit) length(s) real
addresses should be cleaved into a ``public,'' suitably anonymous prefix to be
preserved and reported as is and a private suffix to be discarded
or obscured.  Note that prefix preservation in truncation-based
anonymization, differs from ``prefix-preserving anon\-ymiz\-ation'' in
the literature~\cite{DBLP:conf/imw/XuFAM01,DBLP:journals/cn/FanXAM04}
which preserves prefix {\em lengths} amongst anonymized addresses
but not the original prefix {\em value.} To tackle the problem
of determining whether truncated prefixes or aggregates might
effectively provide anonymity, {\em i.e.,} to make an individual appear
indistinguishable amongst a set of individuals
(see Section 6.1.1.~\cite{RFC6973}),
we count active addresses to
determine how many they actually aggregate. Then, we use
such counting as the basis for anonymization by {\em variable length}
truncation or aggregation, resulting in different lengths to
anonymize different areas of the address space.

The key to our technique is to count a subset of simultaneously
assigned, active IPv6 addresses given the likelihood that a given
temporary privacy address must still be assigned in between two times at
which the address' activity was observed from a (possibly
remote) vantage point.  We rely on the
ostensibly unique identifier present in temporary addresses that
employ Stateless Address Autoconfiguration (SLAAC) with privacy
extensions~\cite{rfc4941}.  Far and away, this is the most common
address assignment mechanism for World-Wide Web (WWW) clients. As of
March 2017 by the 3-day stable definition, {\em i.e.,} ``3d-stable
(-7d,+7d)''~\cite{plonka2015temporal}, we find 684 million
(93\%) ephemeral IPv6 WWW client addresses are active per day and
4.33 billion (97\%) per week.  While temporary privacy
addressing aims to improve privacy by complicating the tracking of user
activity beyond hours or days, we manage to use logs of sporadic
activity of these short-lived addresses to calculate a lower bound
on the number of simultaneously-{\em assigned} IPv6 addresses in a
given prefix {\em even during times when those addresses' seem
inactive}, {\em e.g.,} from a remote CDN vantage point.

To the best of our knowledge, prior works have not reported the
privacy concerns we have with IPv6 address anonymization by truncation nor
have they proposed privacy guarantees with such methods.
Our goal is to develop IPv6 address anonymization that yields
precise, useful results for operations and research while
guaranteeing address privacy for users.  Although this is a work
in progress, we offer the following contributions: $(i)$
an evaluation of IPv6 address anonymization by truncation; $(ii)$
our {\em kIP anonymization} method
(inspired by $k$-anonymity~\cite{Samarati98protectingprivacy})
and early results on its performance.

\section{Methods}
\label{sec:method}

kIP anonymization involves three operations, namely:
address classification,
address activity matrix analysis, and
anonymous aggregate (prefix) synthesis.

\subsection{Address Classification}

We employ address classification to identify SLAAC privacy
addresses, {\em i.e.,} those having pseudorandom values in their
64-bit IID.  To do so, every input address is preprocessed
by the {\tt addr6} tool which performs an initial stateless
classification~\cite{ipv6toolkit}.  For example, consider
the 16 IPv6 addresses in Figure~\ref{fig:firstmatrix}; {\tt
addr6} reports each of them as having a {\tt randomized} IID
because they do not have some other easily recognized IID type,
{\em e.g.,} EUI-64, nor an easily recognized pattern, {\em e.g.,}
only low-bytes being non-zero~\cite{gont2016network}. Next,
we perform a stateful classification using our {\tt dendracron}
tool~\cite{plonka2015temporal,dendrachronology}. By ``stateful,'' we
mean that we classify each address in: $(a)$~{\em space}, relative
to others addresses in a set, {\em e.g.,} those within the same /64
prefix, and in $(b)$~{\em time}, throughout an observation timeframe,
{\em e.g.,}
a week. This yields two classification metrics for each address
that we will use below: {\em (1)} its Discriminating Prefix Length
(DPL) and {\em (2)} its number of Stable Days (SD) during which
we've observed that address to be active and throughout which the
address might have remained assigned.  The DPL is simply the position
of the first
(left-most) bit at which the address differs from its nearest
(observed) address.  The SD is the number of days across which
the address has been seen active.  The smaller an address' DPL
(and SD) value, the more likely it is to have a randomized IID
(and a temporary one at that) by the following rationale.

\textbf{\em Identifying Plausible Randomness:}
Given a set of addresses in a /64 prefix,
the following test for randomness in IIDs complements those above;
it is based 
on the likelihood that a subset of bits at a given position
is distinct across all of the IIDs
assuming the bits were chosen randomly.
For example, suppose there are 2 addresses, and consider the leading 6 bits
of their IIDs.
If these 6 bits were chosen randomly~\cite{rfc4941}, then, out of the $2^6 = 64$ possible 6-bit strings, the likelihood that these two IIDs have different values
for these 6 bits is $63/64 \approx 0.98$.
More generally, given $A$ addresses with candidate random IIDs
in a /64 prefix, and a bit string of length $N$ at a given position in the IIDs,
then the number of possible bit strings is $S = 2^N$, and the probability that 
the bit strings in those IIDs are distinct is:
\begin{equation*}
\frac{S}{S} ~*~ \frac{S-1}{S} ~*~ \frac{S-2}{S} ~*~ \cdots ~*~ \frac{S-(A-1)}{S} 
\end{equation*}
For classification, we start with the number of addresses and a desired probability, {\em i.e.,} 0.99, and compute the number of bits.
In particular, given $A$ addresses with candidate random IIDs, we compute the smallest number of bits, $N$, such that 
the probability is at least 0.99 that all $N$-bit strings at a given position in the IIDs are distinct.
Given this $N$, we examine the IIDs in the /64 to see whether the given bit strings 
are all distinct.
If all bit strings are distinct, then we infer that there is further evidence
that the IIDs are pseudorandom.  If they are not all distinct, then the IIDs may still be pseudorandom, but we choose to not make the inference.
Lets consider the candidate random bit string to begin at the first bit
of the IID.  RFC4941~\cite{rfc4941} dictates that the 7th bit
be set to zero in an otherwise randomized IID,
so, if the bit string spans that bit, we need to allow an additional DPL bit.
Conveniently, this makes $64+1+N$ the 99\%-probable 
{\em maximum} DPL of each address in a set of addresses, of size $A$, having candidate random IIDs.
We implement this additional test for randomness using a precomputed lookup
table in {\tt dendracron}, for $A$ ranging from $2$ to about $1$ million.
For example, for 16 addresses such as those in Figure~\ref{fig:firstmatrix},
the 99\%-likely maximum DPL is $64+1+14=79$; since all the actual DPL values (in the
second column of the figure) are less than $79$, we classify these addresses
as having plausibly random IIDs as a basis for assignment inferences, next.

\subsection{Address Activity Matrix Analysis}

Richter {\em et al.}~\cite{IMC2016Beyond} employed an IP address
activity matrix, with time on the horizontal axis and address space
on the vertical axis, to visualize daily activity over months and to
calculate IPv4 address space utilization.  At finer timescales, we
use an enhanced address matrix to handle the sheer size and sparsity
resulting from IPv6 assignment practices, {\em e.g.,} SLAAC with
privacy extensions.  Figure~\ref{fig:firstmatrix} is an activity
matrix capturing activity per hour for 16 active IPv6 addresses, sorted
by address value, in
one /64 prefix during 24 hours.  An address is active sometime during
each hour marked with ``{\tt \#}'' and inactive other hours from the CDN's
vantage; see legend in
Figure~\ref{fig:matrixlegend}.  The address matrix has temporal
parameters that allow the analysis and method to operate on other
timescales: $i$, the time interval used to aggregate activity and $w$
the time window of observation. Here $i=1$ and $w=24$ in hour units.

Figure~\ref{fig:secondmatrix} is a resorted activity matrix that
contains the same activity information, but with the address' rows
in order of initial activity time (earliest first) rather than by
address value.  This makes clearer which of the addresses {\em might}
be active simultaneously. (Also, when the IID is random,
sorting by address value is meaningless.) We see, for instance,
that two addresses were both active in interval (hour) 3, but we
can not conclude that they were active {\em simultaneously,} since 
interval-binned summaries {\em e.g.,} hourly,
do not record durations of transactions.

In Figure~\ref{fig:thirdmatrix}, we rewrite
Figure~\ref{fig:secondmatrix}'s activity in four ways: address'
activity in just one interval $i$ (within window $w$) is marked
``{\tt X}''; address' activity in multiple, contiguous intervals
have the first interval marked ``{\tt >}'' and the last ``{\tt
<}''; intervals between those at which we {\em infer} that the given
temporary privacy address is {\em assigned} throughout and are marked
``{\tt @}''.

\textbf{\em Address Assignment Inference:}
The ability to infer address assignment between moments of activity
is the key to our method.  Critically, we assume that IPv6
host implementations that support privacy extensions choose good
pseudorandom values when building their IIDs. This allows one to
{\em infer} that a given host's temporary SLAAC address with randomized
IID must still be assigned between (any) two instances of observed
activity since it is ridiculously unlikely that host or any other
will choose the
same pseudorandom value for the IID on any subsequent reconfiguration
of its network interface(s).

Now that we have inferred the intervals in which each address is
assigned, we can count the simultaneously assigned addresses. To do
so, we perform special arithmetic to ``sum'' the marks in each
hour interval (column):
{\bf{\em (1)}}
Each of a column's ``{\tt @}'' marks increment its total by 1; this
is because those addresses were assigned at every moment during
the interval and the moments between the previous and next intervals.
{\bf{\em (2)}}
Either $(a)$, each of a column's ``{\tt >}'' marks can increment its
total by 1 because we know that the moment between this interval and
the next had that additional address assigned; or $(b)$ each of a
column's ``{\tt <}'' marks can increment its total by 1 because we
know that the moment between this interval and the next had that
additional address assigned. Of $(a)$ or $(b)$, we choose whichever
column total would be larger.
{\bf{\em (3)}}
All of a column's ``{\tt X}'' marks, taken together, increment its
total by 1 only in column's having no ``{\tt >}'' or ``{\tt <}'' marks;
this is because we know there was some specific moment amidst that
interval when (at least) one of those addresses was assigned that
wasn't a moment between the previous and next intervals.  This
process is performed for each column to come to a sum, for example,
shown totaled below the matrix in Figure~\ref{fig:thirdmatrix}.
These totals, single digits with whitespace removed here: ``{\tt
000100011112332321122100}'', are the lower bounds on the number of
simultaneously assigned addresses in each interval (hour). Their
minimum is $0$ and maximum is $3$, meaning we are
confident at least 3 addresses were simultaneously assigned within
this /64 prefix at some moment during this day.

The last step in our matrix analysis is to infer precise
moments that the /64 prefix,
itself: {\tt 2001:db8::/64}, {\em must} be assigned to some hosts'
interface, given when the addresses it covers were known assigned.  We do
so by inferring address assignment at ``fenceposts,'' {\em i.e.,}
the moments between our ``fence sections'' (intervals) in time.
Because these are the moments between intervals, they number 1
fewer than intervals in the window: $f=w-1=23$. The /64 prefix
is inferred to have been assigned at the fencepost trailing each
interval where there is a {\tt !} (exclamation point) mark.
We now have a time series array of size $f$, temporarily encoded:
``{\tt --------!!!!!!!!-!!!!--?},'' that indicates the precise moments
(between intervals) when the prefix must have been assigned.
Translating the {\tt !} marks to $1$ (and others to $0$) makes them
time series values suitable for accumulating across an entire network
(or the entire Internet) to compute $f$ (hourly) lower bounds on the count of
simultaneously assigned /64 prefixes.

\lstset{keepspaces=true,basicstyle=\scriptsize\ttfamily}

\newsavebox{\firstmatrix}
\begin{lrbox}{\firstmatrix}
\begin{lstlisting}
                              D 
                              P  S       hour of day
IPv6 address                  L  D 0         1         2
----------------------------- -- - 012345678901234567890123
2001:db8::117a:e091:b2bd:ca65 67 0 |-------+-------+--##---
2001:db8::21ad:6d24:641a:1314 68 0 |--#----+-------+-------
2001:db8::3454:ae0d:20a0:df4d 68 0 |-------+--#----+-------
2001:db8::4974:fa8b:465d:4c2a 68 0 |-------+-------+#---#--
2001:db8::503c:a91d:be00:9a63 68 0 |-------##-###--+-------
2001:db8::6867:8a64:5417:e731 70 0 |-------+---##--+-------
2001:db8::6d35:ee11:ec45:f658 70 0 |-------+-------+#------
2001:db8::7070:a7fc:47d5:02ba 70 0 |------#+-------+-------
2001:db8::7554:b66a:a983:9665 70 0 |-------+--#----+-------
2001:db8::7939:1bd6:fec2:85bb 70 0 |-------+------#+-------
2001:db8::7ccc:3977:7c76:bdef 70 0 |-------+-------+---#---
2001:db8::890b:1f0d:14e2:0ccb 67 0 |-------+----#--+-------
2001:db8::a0fc:1e18:48aa:eb2e 67 0 |-------+---#---#-------
2001:db8::f930:9833:f8c5:3926 74 0 |-------+----#--#-------
2001:db8::f94d:fcec:6b8e:d61f 74 0 |-------#-------+-------
2001:db8::fd28:50fe:8445:83e7 70 0 |--#----+-------+-------
2001:db8::/64 16; Temporary SLAAC: 100.00%

2001:db8::/64 -------------------------------------------->
\end{lstlisting}
\end{lrbox}

\newsavebox{\secondmatrix}
\begin{lrbox}{\secondmatrix}
\begin{lstlisting}

       hour of day
 0         1         2
 012345678901234567890123
 |--#----+-------+-------
 |--#----+-------+-------
 |------#+-------+-------
 |-------##-###--+-------
 |-------#-------+-------
 |-------+--#----+-------
 |-------+--#----+-------
 |-------+---##--+-------
 |-------+---#---#-------
 |-------+----#--#-------
 |-------+----#--+-------
 |-------+------#+-------
 |-------+-------+#---#--
 |-------+-------+#------
 |-------+-------+--##---
 |-------+-------+---#---


 ----------------------->
\end{lstlisting}
\end{lrbox}

\newsavebox{\thirdmatrix}
\begin{lrbox}{\thirdmatrix}
\begin{lstlisting}

        hour of day
  0         1         2
  012345678901234567890123
  |--X----+-------+-------
  |--X----+-------+-------
  |------X+-------+-------
  |------->@@@@<--+-------
  |-------X-------+-------
  |-------+--X----+-------
  |-------+--X----+-------
  |-------+---><--+-------
  |-------+--->@@@<-------
  |-------+---->@@<-------
  |-------+----X--+-------
  |-------+------X+-------
  |-------+-------+>@@@<--
  |-------+-------+X------
  |-------+-------+--><---
  |-------+-------+---X---
  000100011112332321122100
  |-------+-------+-------
  --------!!!!!!!!-!!!!--?
\end{lstlisting}
\end{lrbox}

\newsavebox{\legend}
\begin{lrbox}{\legend}
\begin{lstlisting}
| = every 24th interval in the matrix, e.g., first hour of day, UTC
+ = every 8th interval in the matrix, e.g., 8 hours
# = activity logged during the given hour

X = activity started and ended during the given hour, i.e., a "short" episode
> = activity started during the given hour, i.e., beginning of an episode
< = activity ended during the given hour, i.e., end of an episode
@ = infer address assigned throughout the given interval, e.g., hour
! = infer /64 prefix assigned at trailing edge of given hour, i.e., the "fencepost" moments between intervals
? = the last "fencepost" moment is discarded since address assignment can't be determined (yet)
\end{lstlisting}
\end{lrbox}

\begin{figure*}
\begin{center}
\subfigure[Initial Activity Matrix: 16 addresses in 1 /64 prefix]{\label{fig:firstmatrix}\usebox{\firstmatrix}}
\subfigure[Time Sorted Activity]{\label{fig:secondmatrix}\usebox{\secondmatrix}}
\subfigure[Inferred Assignment]{\label{fig:thirdmatrix}\usebox{\thirdmatrix}}
\subfigure[matrix legend]{\label{fig:matrixlegend}\usebox{\legend}}
\vspace{-3mm}
\caption{$(a)$ Address classifications (DPL, SD) and activity matrix for one /64 prefix: \texttt{2001:db8::/64}, here, having 16 SLAAC addresses observed as active during 24 hours time.
In the matrices, space is represented vertically and time progresses horizontally, left to right.
The matrix' addresses are then sorted $(b)$ by their initial activity times. Finally, we infer $(c)$ the number of simultaneously assigned IPv6 addresses and /64 prefix(es) on-off times.
\label{fig:matrix}}
\end{center}
\end{figure*}

\newsavebox{\meetingmatrix}
\begin{lrbox}{\meetingmatrix}
\begin{lstlisting}
                   /64 Inferred Assignment Matrix
                         ------------------------
                               hour of day
                         0         1         2
                         012345678901234567890123
2001:db8:370::/64        !!!!!!!!!!!!!!!!!!!!!!! 
2001:db8:370:128::/64    !!!!!!!!!!!!!!!!!!!!!!! 
2001:db8:370:228::/64    !------------!--------- 
\end{lstlisting}
\end{lrbox}

\begin{figure*}
\begin{center}
\subfigure[]{
\label{fig:meetingtree}
\includegraphics[width=0.75\textwidth]{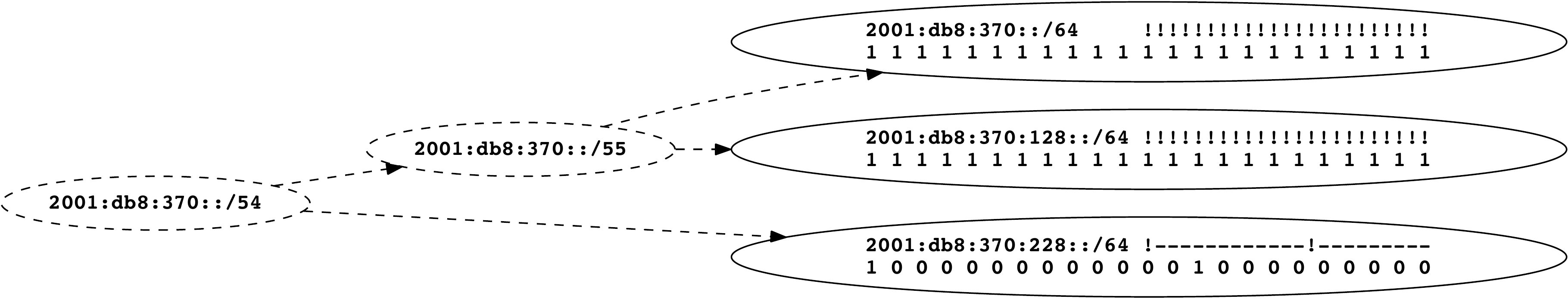}
}
\subfigure[]{
\label{fig:meetingtreetoo}
\includegraphics[width=1.0\textwidth]{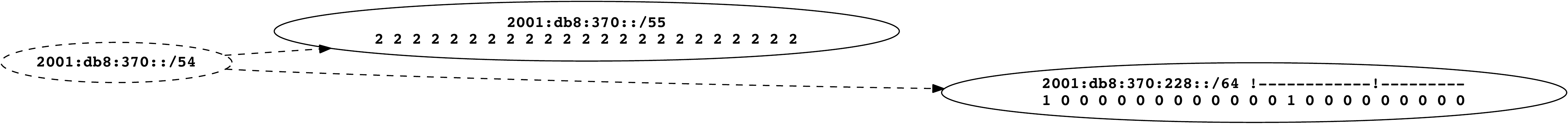}
}
\subfigure[]{
\label{fig:meetinganon}
\includegraphics[width=0.8\textwidth]{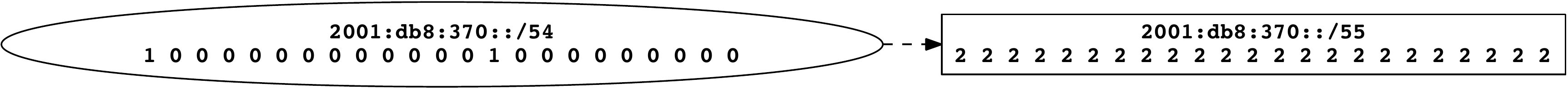}
}
\vspace{-3mm}
\caption{
A 3-step method to calculate anonymous aggregates
using a binary PATRICIA tree, for example, the
Meeting Network in Table~\ref{tab:dataSummary}. (a) First, a tree is
populated with active /64 prefixes (the 3 nodes with solid lines), each
with a time series showing assignment, on or off. (b) Next, the tree is
aggregated and the arrays added to their parent until the min count
of $k=2$ is reached. Here, the first two /64s have been aggregated to their
parent, but the third has not yet been visited. (c) Finally, after
aggregation is complete, we report only the anonymous aggregates
with (at least) the desired number of simultaneous addresses ($2$):
only the one prefix shown in a solid box.
\label{fig:meeting}}
\end{center}
\end{figure*}

\subsection{Synthesizing Anonymous Aggregates}

Our synthesis of aggregates, {\em i.e.,} larger, covering prefixes
that coalesce individuals' SLAAC prefixes for privacy, is inspired by
Cho {\em et al.}~\cite{DBLP:conf/qofis/ChoKK01,plonka2015temporal}
who developed ``aguri'' which recursively aggregates prefixes based
on activity counts until a stopping condition is met.  We augment
the aguri tree with an array (size $f$) of time series counters
in each node and with a new stopping condition where a statistical
function of those arrays' (hourly) values, {\em i.e.,} either minimum
(min) or maximum (max) meets or exceed $k$, the desired number of
active addresses that can not be distinguished once anon\-ym\-ized.
We use min and max to show the range of outcomes, and choose median
as a robust metric to avoid outliers, {\em e.g.} due to flash crowds,
quiescent hours of day, or sustained outages.

Figure~\ref{fig:meeting} shows a simple example for active
address in a Meeting Network, which had only 3 active /64 prefixes covering
all the attendees active addresses (as shown in Table~\ref{tab:dataSummary}),
despite the meeting running for a week and providing IPv6 connectivity,
both wired and wireless, to its $1000+$ attendees.
With minimum $k=2$, our method reports one anonymous aggregate,
\texttt{2001:db8:370::/55}, because this network has few active /64 prefixes.

\begin{table*}[ht!]
\scriptsize
\centering
\begin{tabular}{|l||r|r|r|r|r|r|} \hline
          &                            & {Total} & {Total}        & {\bf Lower bound simultaneously}  & {\bf Lower bound simultaneously} & {Total} \\
{Data set} & {7-day date range} & {active /48} & {active /64}       & {\bf assigned /64 prefixes} & {\bf assigned addresses} & {active} \\
{}        &                            & {prefixes} & {prefixes} & {\bf maximum (median)} & {\bf maximum (median)} & {addresses} \\
\hline\hline
Meeting Network & Mar 25-31 2017 &     1 &     3 &     {\bf 3 (2)} & {\bf 309 (84)} & 15.4K \\ \hline\hline
EU ISP          & Sep 17-23 2016 & 163K & 21.5M & {\bf 2.02M (1.52M)} & {\bf 3.80M (2.63M)} & 125M \\ \hline
JP ISP          & Mar 17-23 2017 & 2.46M & 2.46M & {\bf 1.21M (897K)} & {\bf 2.26M (1.54M)} & 72.2M \\ \hline
US ISP          & Mar 17-23 2017 & 8.16K & 2.42M & {\bf 1.81M (1.66M)} & {\bf 4.71M (3.82M)} & 84.5M \\ \hline
\end{tabular}
\caption{Characteristics of the active IPv6 WWW client address data sets.
Counts determined by our method, as described in Section~\ref{sec:method}, are shown {\bf bold}. The others are simple activity counts during the 7 days of observation.
\label{tab:dataSummary}}
\end{table*}

\section{Results}

We offer two early results based on logged IPv6
addresses of active WWW clients from real networks: {\em
(1)}~an evaluation of current practice for truncation-based
anony\-miz\-ation and {\em (2)}~a characterization of the anonymous aggregates
produced by $k$IP anony\-miz\-ation. Both these use real WWW client addresses
found in the activity logs of a large CDN as input data.
Consider Table~\ref{tab:dataSummary}
which characterizes our active WWW client address data sets.
We chose these specific networks for the variety of address assignment
practices they demonstrate, rather than, {\em e.g.,} their country.
We first show the Meeting Network
used in Figure~\ref{fig:meeting} in Section~\ref{sec:method}.
Note the lower bounds for simultaneously assigned /64 prefixes and addresses
and their plausibility with 1000 attendees,
a subset of whose wired and wireless hosts (apparently) use SLAAC.

Table~\ref{tab:dataSummary} also shows three ISP networks,
one each from Europe (EU), Japan (JP), and the United States (US).
Seen over the 7 days of observation,
the three ISPs have wildly varying numbers of active /64 prefixes
(up to $\sim 10\times$ different) and /48 covering prefixes
(up to $\sim 300\times$ different).  This strongly suggests that either
{\em (a)} their subnet and address assignment practices differ greatly or
{\em (b)} their WWW client population sizes differ greatly.
However, note that they share similar lower bounds on number of
simultaneously assigned /64 (SLAAC) prefixes and addresses calculated
by our method, each being in the low single-digit millions.
These intermediate results, {\em i.e.,} estimated counts of
each network's IPv6-capable WWW clients, suggests that our method
can mitigate bias in counting caused by some networks' address
assignment practices and is a basis of ongoing work.  In networks
whose IPv6 addresses' network identifiers (in addition to IIDs) contain
pseudorandom segments or have short assignment periods from very large
pools, simple counts, {\em e.g.,} active prefixes per day or week,
can far exceed the number of IPv6-capable clients that actually
exist in a given network. Thus, more careful counting is warranted
as the basis for guaranteed anonymity in aggregates.

In our results below, we used the full ISP data sets in
Table~\ref{tab:dataSummary}; meaning $w=168$ (overall window
of time) and $f=167$ (fencepost moments) in units of $i=1$ hour
(intervals). Thus, min is the minimum
of the 167 lower bounds counts of simultaneously-assigned /64
prefixes, {\em i.e.,} 167 hours (one week's time). Likewise,
max and median are of the 167 (hourly) lower bound counts.

\subsection{Evaluation of /48 Aggregation}

Today, a not uncommon practice for IPv6 address anonymization
is to truncate the low 80 bits of the address, preserving only the
address' /48 (covering) prefix as the supposed anonymous aggregate.
Based simply on counting, in Table~\ref{tab:dataSummary}, we see
that the JP ISP has the same count (2.46 million) of active /64
prefixes and /48 prefixes.  This means, by and large, the JP ISP does
not use the bits 48-63 to differentiate individual SLAAC-prefixed
subscribers. (Those bits happen to be zeroes, but they could use
any value to the same effect.)  Thus, truncating the last 80 bits
of the JP ISP addresses does nothing to obscure the SLAAC-prefix
ostensibly associated with an individual customer. This proves 80
bit truncation, and /48 aggregation, is ineffective at anonymizing
a customer's SLAAC prefix in this network.

\begin{figure}[h!]
\centering
\includegraphics[scale=0.34]{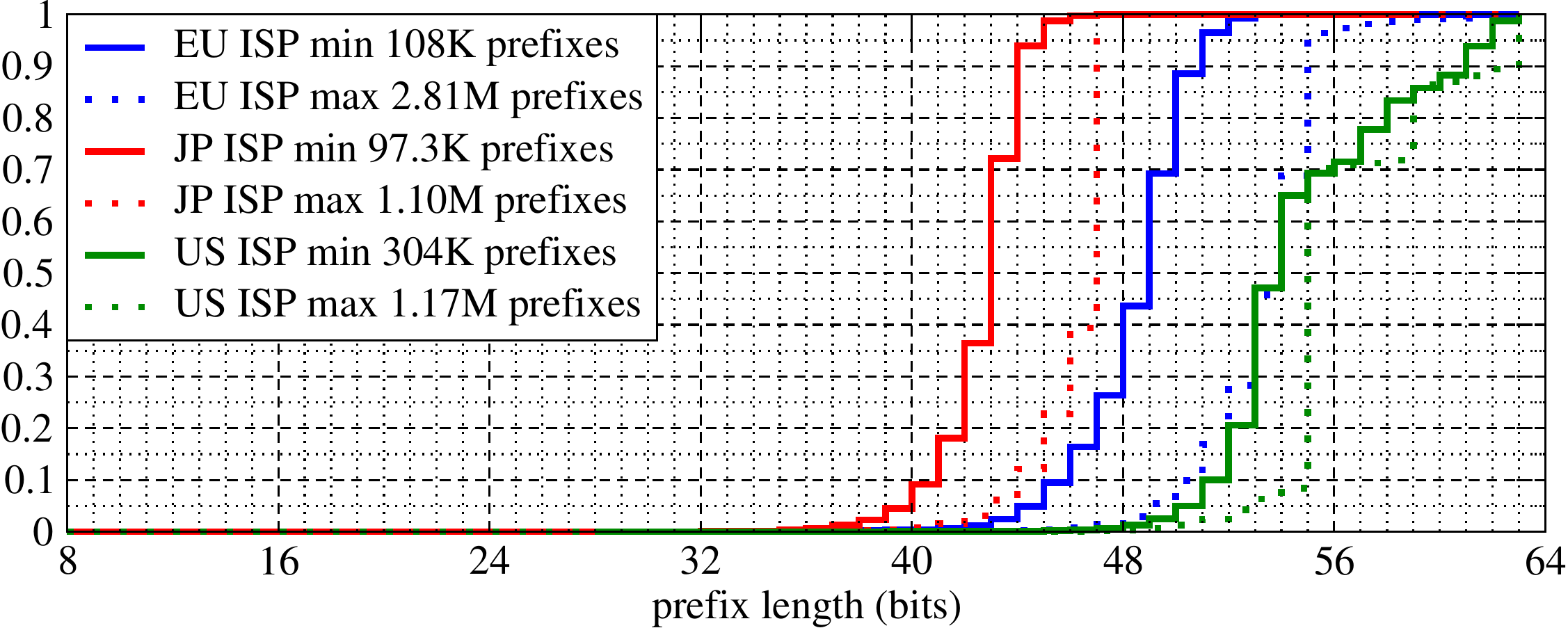}
\caption{CDF of aggregate prefix lengths, $k=2$.} \label{fig:k2} 
\end{figure}

To explore this more broadly, Figure~\ref{fig:k2} plots the
distribution of prefix lengths necessary to aggregate 2 simultaneously
assigned active /64 prefixes together.
(IPv6 hosts have a 64-bit subnet prefix~\cite{RFC4291,RFC7421},
thus an ISP commonly provides at least a /64 prefix to each customer
or subscriber.)
This is based on our 
empirical measurements and lower bounds on numbers of
simultaneously-assigned IPv6 /64 prefixes as described in
Section~\ref{sec:method}.
In this figure, note the JP ISP's (red) $k=2$  min and max prefix lengths
always plot left of bit 48 on the horizontal axis; as above, this confirms
that aggregation to /48 would {\em never} aggregate any individual
customer's SLAAC prefixes together, thus truncation to /48 does not
improve privacy to subscribers.  The figure also shows (solid
blue) that the EU ISP's $k=2$ min more than 25\% of the prefixes
needed prefix lengths less than 48, {\em i.e.,} not even two /64s are
aggregated by /48s therein.
Moreover, we presume no one would settle for such a weak notion
of anonymity as $k=2$, which would be equivalent to truncating only 1
bit in IPv4, or aggregating to /31. Both in the past and today with IPv4,
it is common to truncate or aggregate to /24 prefixes, presumably with
the intention of aggregating up to 256 (max) individuals' addresses
together. We consider more reasonable values of $k$ for privacy, next.

\subsection{Anonymous Aggregates}

\begin{figure}[h!]
\centering
    \subfigure[histogram: median $k=32$]{\includegraphics[scale=0.34]{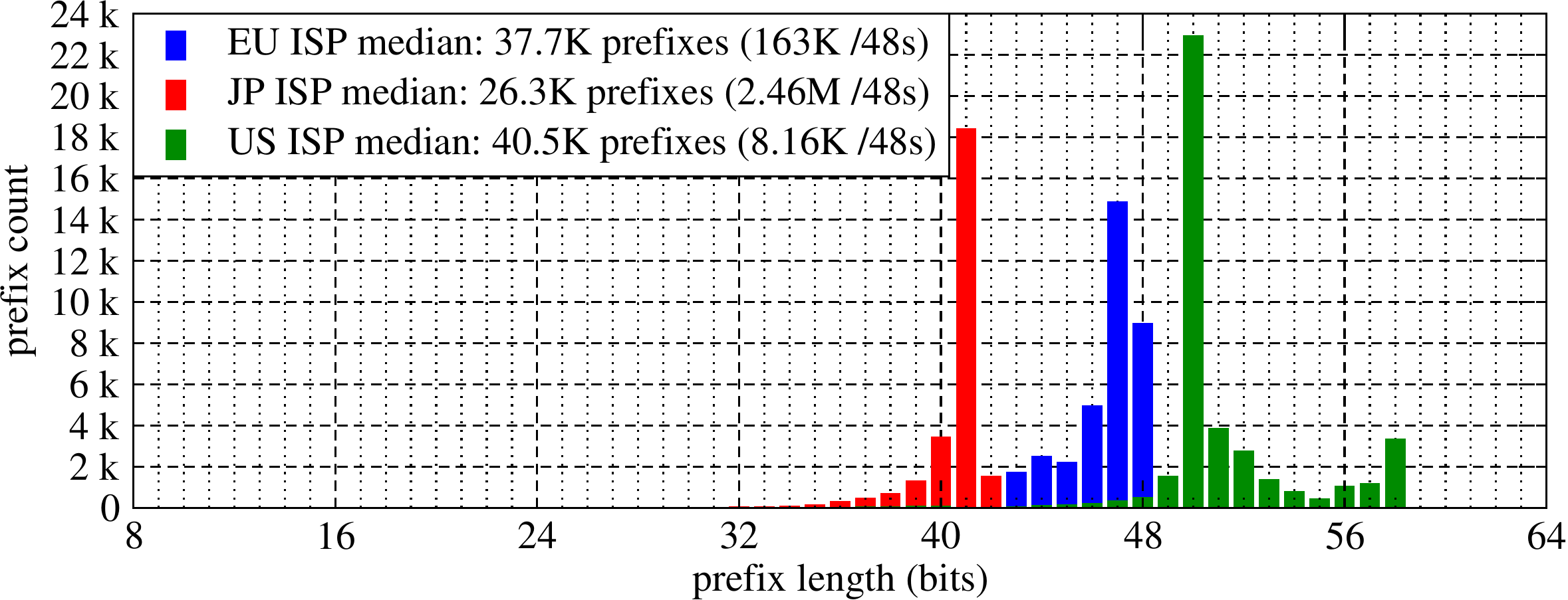} \label{fig:k32hist}}\\
    \subfigure[histogram: median $k=256$]{\includegraphics[scale=0.34]{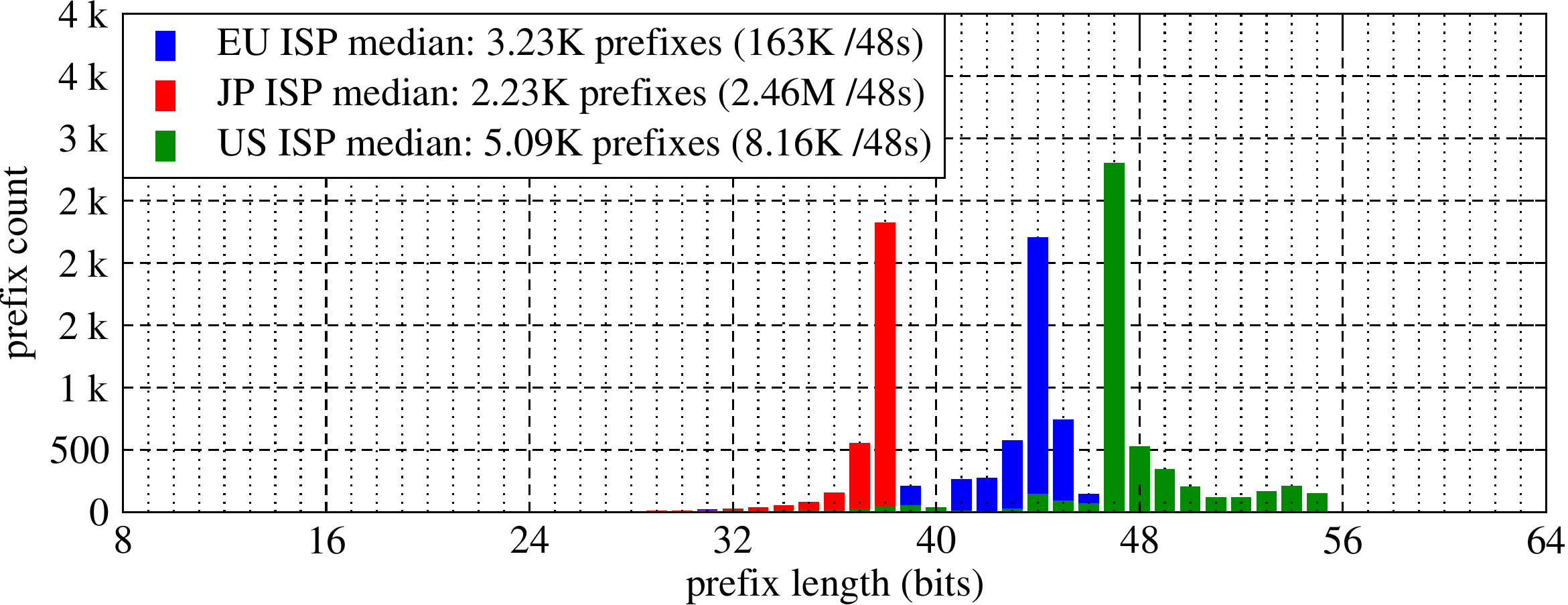} \label{fig:k256hist}}
\caption{Anonymous aggregate prefix lengths.} \label{fig:aggCDF} 
\end{figure}

We now apply $k$IP anon\-ymization to addresses for each of the
three ISPs.  Figure~\ref{fig:aggCDF} characterizes the resulting
prefixes in histograms for $k=32$ and $k=256$. (Essentially, these
are equivalent to aggregating a fully-utilized IPv4 network to /27 or
/24, respectively.)  At these levels of anonymization, we find that,
the JP ISP and EU ISP almost always required more aggregation
than /48 (more than 80 bits truncated) for us to {\em guarantee}
that the aggregation meets our desired $k$ on lower bounds for
median counts of simultaneously-assigned addresses. The $k$IP
anonymous aggregate prefixes reported here vary from /25 to /58.
In CDF plots (not shown) for the US ISP's customer's, we find
that /48 aggregation guarantees $k=32$ anonymization for 90-95\%
(min-max) of those customers, but guarantees $k=256$ anonymization
for only 30-40\% of those customers.  By comparing the resulting
$k$IP anonymous aggregate prefixes counts to fixed-length /48
prefix counts, as shown parenthetically in the legends in Figure~\ref{fig:aggCDF},
by and large, we see that $k$IP anonymization (with $k=256$ or even
$k=32$) can yield a much smaller set of anonymous prefixes while
guaranteeing significant aggregation of individuals' /64 prefixes.

\vspace{5mm}
\section{Limitations \& Future Work}

In offline operation, {\em e.g.,} anonymizing addresses of clients
that access a worldwide CDN as we report here, the guarantee of
anonymization is strong because every address observed is used as
input to the anonymization. However, in online operation, {\em e.g.,}
anonymizing addresses on the fly, the guarantee is unclear because
the anonymous aggregate set was determined {\em a priori,}
based on addresses active in the past. Thus, online operation of
this method entails {\em forecasting}, wherein the anonymizer likely
assumes that past activity is suitably representative.

Lets consider attacks and situations that might
call our claimed privacy guarantee into question. 
In this work, we treat an address' /64 prefix and anything more specific,
{\em e.g.,} the IID, as private.
While it's common for ISPs to provide a /64 prefix to a
customer, some ISPs will honor requests for a larger prefix,
{\em e.g.,} a /60 or /56~\cite{RFC7550,JohnB,LeeH}. Then,
the customer's router can advertise SLAAC prefix(es) to the their hosts.
In this case, it is possible for an individual customer to have a
set of simultaneously-assigned /64 prefixes, resulting in an
anonymous aggregate where the number of distinct {\em customers}
could be much less than $k$.
To combat this, an anonymizer wants to know the customer's prefix length,
so that it might increase $k$ accordingly.  Discovering this prefix
length automatically (via the activity matrix) is ongoing work.
Similarly, if a malicious party generates traffic
from what would be quiescent source addresses in many unique /64 prefixes, they
might cause $k$IP-anonymization to report more specific anonymous
aggregates allowing them to determine what their neighbors' nearest
active prefixes might be.  For this reason, it may be important to keep time
series of simultaneously assigned address counts (as we do here),
so that anomalous counts, {\em e.g.,} during flash crowds, or
attacks, can be identified and/or ignored.

In conclusion, we evaluate IPv6 address anon\-ym\-ization and
demonstrate that truncation to a single prefix length of 48 bits,
Internet-wide, fails to anonymize information associated with
individuals' IP address identities, {\em e.g.,} /64 prefixes, in
the face of some common addressing practices used today.  We develop a
technique to compute lower bound counts on simultaneously-assigned
addresses and an improved anonymization method that truncates to
variable prefix lengths guaranteeing a desired degree of address
privacy.  Our results show that $k$IP anonymization, {\em e.g.,} with
$k=32$, outperforms IPv6 address anonymization by 80-bit truncation.
Thus, we propose this as preferred privacy practice in research and
operations and invite community feedback.

\section*{Acknowledgments}

We thank David Duff, KC Ng, Ramakrishna Padmanabhan,
and Paul Saab for their assistance and helpful comments.

\balance
\bibliographystyle{plain} 
\bibliography{main}

\end{document}